# Fe implanted ferromagnetic ZnO

K. Potzger, Shengqiang Zhou, H. Reuther, A. Mücklich, F. Eichhorn, N. Schell, W. Skorupa, M. Helm and J. Fassbender

Institute of Ion Beam Physics and Materials Research, Forschungszentrum Rossendorf, P.O. Box 510119, 01314 Dresden, Germany

T. Herrmannsdörfer, T. P. Papageorgiou

Hochfeld-Magnetlabor Dresden, Forschungszentrum Rossendorf, P.O. Box 510119, 01314 Dresden, Germany

Room-temperature ferromagnetism has been induced within ZnO single crystals by implant-doping with Fe ions. For an implantation temperature of 620 K and an ion fluence of $4\times10^{16}$ cm$^{-2}$, very tiny Fe particles, formed inside the host matrix, are responsible for the ferromagnetic properties. They were identified using synchrotron X-ray diffraction and Mössbauer spectroscopy. On the other hand, Fe ions implanted at a temperature of 253 K and an ion fluence of $4\times10^{15}$ cm$^{-2}$ are incorporated into the host matrix and develop a room temperature diluted magnetic semiconductor (DMS).

In the field of spintronics[1], diluted magnetic semiconductors (DMS) are worldwide under intense investigation. DMS are "conventional" semiconductors doped with transition metal (TM) or rare-earth ions which are diluted within the host matrix and ferromagnetically aligned via an indirect magnetic coupling[2-7]. The existence of DMS basing on Mn doped p-type ZnO[2] and V, Ti, Fe, Co or Ni doped n-type ZnO[7] has been predicted by theory. However, currently only n-type conducting ZnO films or single crystals are available. Recent reviews of experimental work on the field, is given by S. J. Pearton[8] *et al.* and Ü. Özgür *et al.*[9] Among other systems, n-type ZnO doped with Fe has been confirmed experimentally[10-12] to exhibit ferromagnetism at room temperature. In some cases, especially at high processing temperatures, unwanted secondary phases are formed inside the ZnO matrix, which are responsible for the ferromagnetic properties[11]. One way to overcome this problem is the use of ion beam doping at low temperatures and thus far from thermal equilibrium[12-14]. In any case, structural analysis methods with high sensitivity are necessary in order to exclude secondary phases. In this letter it will be shown that Fe-implantation into ZnO single crystals at a temperature of 620 K can lead to the formation of ferromagnetic α-Fe nanoparticles. On the other hand, Fe ions implanted at a temperature of 253 K are diluted within the ZnO host matrix and develop a ferromagnetic coupling.

For this purpose we used commercially available, hydrothermally grown ZnO single crystals that have been Zn-face epi-polished by the supplier. These samples were implanted with $^{57}$Fe ions at different temperatures and ion fluences (for a sample register and abbreviations see Table I). The implantation energy of 180 keV yielded a projected range of $R_P$=83±35 nm (TRIM code[15]). Prior to implantation, the virgin samples were characterized by X-ray diffraction (XRD, Siemens D5005), inductive coupled plasma mass spectrometry (ICPMS), and superconducting quantum interference device (SQUID, Quantum Design MPMS) magnetometry. It is found that the virgin crystals are perfectly single crystalline showing a contamination below 20 ppm for Cu, Ni and Fe and below 0.1 ppm for the other transition metals. Most important is the fact that even at low temperatures (5 K) all of the virgin samples behave purely diamagnetic upon magnetization reversal.

After implantation, the four samples (Table I) were analyzed using SQUID magnetometry. It was found that only two of them, i.e. the HFHT and the LFLT samples exhibit a pronounced hysteresis loop upon magnetization reversal at T=5 K (Table II). After subtraction of the diamagnetic background, a saturation magnetization of $M_S$=0.30 $\mu_B$ ($M_S$=1.3 $\mu_B$) per implanted Fe ion and a coercivity of $H_C$=2.4×10$^4$ Am$^{-1}$ ($H_C$=4.8×10$^3$ Am$^{-1}$) for the HFHT (LFLT) sample is determined. The hysteretic behavior remains also at T=300 K (Fig. 1a, b). However, for the HFHT sample a more drastic decrease of $M_S$ and $H_C$ as compared to the LFLT sample is observed with increasing temperature (Table II).

In order to analyze the microscopic origin of the measured ferromagnetic properties of the HFHT- and LFLT-samples, synchrotron X-ray diffraction (SR-XRD) with monochromatic X-rays of 0.154 nm wavelength and room-temperature conversion electron Mössbauer spectroscopy (CEMS) were used for all samples. In contrast to conventional XRD, the much higher X-ray intensity in SR-XRD allows one to detect also small amounts of very tiny nanoparticles. Fig. 1c shows a symmetric 2θ/ω scan for the HFHT sample. Sharp, high intensity peaks from bulk ZnO are visible at 2θ ~ 34.4° and 2θ ~ 72.6°. At 2θ ~ 44.5°, a rather broad and low intensity peak originating from α-Fe(110) with a theoretical Bragg angle of 2θ=44.66° occurs. The nanoparticle size is estimated to be around 8 nm using the Scherrer formula[16]. Apart from α-Fe, no other phases are detected. In order to support these findings by real space methods, cross-section transmission electron microscopy (TEM, Philips CM 300) has been performed. The nanoparticles could be identified indirectly due to a Moiré - pattern with a visible diameter of about 6 - 12 nm (Fig. 2, inset) at a distance of only 30-65 nm from the surface. Thus, nanoparticle formation preferentially occurs close to the surface, considering the calculated $R_P$ of 83 nm with a straggling of ±35 nm. A broadening of the Fe density profile due to the elevated implantation temperature was observed by means of energy dispersive X-ray microanalysis (EDX). The maximum iron concentration was found at $R_{EDX}$=75±55 nm. In CEMS, only the HFHT sample exhibits a fraction of $^{57}$Fe probe nuclei, that show a clear magnetic hyperfine splitting (sextet) corresponding to a magnetic hyperfine field of $B_{HF}$=30.5 T (Fig. 2) which is - due to size effects - slightly smaller than the known value of metallic α-Fe ($B_{HF}$=33.0 T). This fraction covers 12.5 % of the $^{57}$Fe nuclei absorbing the incident γ-radiation. Its isomer shift (IS) of 0.06 mm/s with respect to α-Fe doubtlessly represents metallic $Fe^0$. The remaining Fe in the HFHT sample exhibits ionic charge states showing no ferromagnetic hyperfine splitting. The interpretation of these fractions is, in part, rather difficult. The best fit has been obtained using one singlet representing a $Fe^{3+}$-state reported already elsewhere[17] and two quadrupole-split lines representing $Fe^{2+}$ states (Fig. 2). The absence of a quadrupole splitting (QS) of $Fe^{3+}$ excludes $ZnFe_2O_4$-precipitates since there is always an electric field gradient present at the octahedral sites[18,19]. $Fe_3O_4$ usually does not show a quadrupole splitting in the $Fe^{2+}$ states[17] and can thus also be excluded indirectly. These conclusions are consistent with those obtained from SR-XRD.

The interpretation of the origin of the ferromagnetic properties of the HFHT sample is thus straightforward: During implantation metallic Fe-nanoparticles are formed. This is due to higher migration of Fe at the elevated temperature as compared to the HFLT and LFLT samples. Moreover, the required diffusion length for nanoparticle formation is much shorter at the higher fluence as compared to the LFHT sample (Table I). The superparamagnetic limit of Fe nanoparticles is described by the relaxation time $\tau = \tau_0 \exp[\frac{E_A V}{k_B T}]$, where $E_A$ is the anisotropy energy density (5×10$^4$ J/m$^3$ for Fe), V is the particle volume and $k_B$ is the Boltzmann constant. $\tau_0$ amounts to ~$10^{-9}$ s (Ref. 20). Thus at T=5 K and a measurement time of ~100 s which is typical for SQUID magnetometry the critical nanoparticle diameter for superparamagnetic behavior results to 4 nm. From the above discussed structural analysis we know that all nanoparticles diameters are larger than this value and should intrinsically behave like ferromagnetic α-Fe bulk material. Taking into account the fraction of 12.5 % of metallic Fe found by CEMS and a magnetic moment of 0.30 $\mu_B$ per implanted Fe ion, a value of $M_S$=2.4 $\mu_B$ per Fe atom within the metallic nanoparticles is determined in agreement with the known value for bulk Fe of 2.2 $\mu_B$. The slight overestimation probably results from the fact that the CEMS spectrum contains also a small fraction resulting from superparamagnetic Fe nanoparticles which could not be resolved in CEMS. At 300 K, the hysteresis loop obtained by SQUID magnetometry exhibits a distinct decrease of $M_S$ down to 0.17 $\mu_B$ per implanted Fe ion, and of $H_C$ down to $2.4\times10^3$ Am$^{-1}$ compared to the measurement at 5 K (Table II). Both effects result from the size distribution of the Fe-nanoparticles, since with increasing temperature also larger nanoparticles become superparamagnetic or approach to the superparamagnetic limit.

In contrast to the other three samples, a long-time CEMS spectrum (500 hours) recorded for the LFLT sample ($T_{imp}$=253 K, Φ=$4\times10^{15}$ cm$^{-2}$) exhibits only a single line corresponding to a $Fe^{3+}$ state. Thus the majority of the detected ions are ferric but nonmagnetic similar to the results for Fe doped $SnO_2$[5]. A decision about the existence of a ferromagnetic sextet could not be provided along with CEMS due to the small counting rate resulting from the low fluence implanted and the lower uniformity of the Fe lattice sites as compared to the samples implanted at 620 K. However, for the LFHT, HFLT and especially the ferromagnetic LFLT - sample no secondary phases have been found using SR-XRD (Fig. 1d) and no metallic $Fe^0$ states have been detected using CEMS. *Consequently, the implanted Fe-ions are diluted* *within the ZnO host matrix.* Thus - in sharp contrast to the HFHT sample - the ferromagnetic behavior of the LFLT sample (Fig. 1b) results from an indirect exchange interaction between diluted Fe ions similar to the one reported in Ref. 5 for the case of Fe doped $SnO_2$. Surprisingly the ferromagnetic behavior occurs at much lower Fe concentrations than that reported in Ref. 5 or predicted by theory. In the case of diluted $Fe^{3+}$ (5 $\mu_B$ per ion), 28 % and in the case of diluted $Fe^{2+}$ (6 $\mu_B$ per ion), 23 % of the implanted ions would contribute to the ferromagnetic interaction. The minimal Fe-Fe distance for the LFLT sample can be estimated to be 1.3 nm. Considering the different implantation temperatures and fluences affecting the diffusion behavior and the ion induced damage in the four investigated samples, a crude explanation of their behavior with respect to the formation of a DMS can be provided: An implantation temperature of 620 K causes a broadening of the Fe implantation profile. Hence, a ferromagnetic state of the Fe ions that are diluted within the LFHT sample cannot be established due to the low local Fe concentration. Within the LFLT sample however, the implantation profile is sharper and therefore the local Fe concentration is large enough to form a room-temperature DMS. For the lack of a DMS state within the high fluence implanted samples this argumentation does not hold because the total amount of implanted Fe ions was 10 times larger than for the LFHT sample, but with many associated defects. Thus these defects introduced during implantation must play a key role for the DMS formation. Rutherford backscattering (RBS) analysis shows, that the damage level for both high fluence implanted samples are similar, i.e. $\chi_{min}$~65 %[21], while the damage level for the LFLT sample is much lower ($\chi_{min}$=30 %, $\chi_{min}$ of the virgin samples: 3 %). Such defects affect the transport properties of ZnO[22] and thus the path of ferromagnetic coupling.

In summary, 180 keV Fe implanted ZnO single crystals can develop ferromagnetic properties that are either caused by α-Fe nanoparticles or an indirect coupling of the Fe ions in a DMS system, depending on the details of ion fluence and implantation temperature. Detailed structural analysis is required to rule out secondary phases.

**Table captions:**

Table I. Implantation conditions for $^{57}$Fe ions for the investigated samples ($T_{imp}$=implantation temperature, $\Phi$=ion fluence). The implantation angle was set to 7° in order to avoid channeling effects. The calculated implantation profile thus has a Gaussian shape with a maximum atomic concentration $\rho_{max}$ indicated. The sample identifier refers to low/high fluence and low/high temperature.

| Sample | $\Phi$ (cm$^{-2}$) | $\rho_{max}$ (%) | $T_{imp}$ (K) |
|---|---|---|---|
| LFHT | $4\times10^{15}$ | 0.5 | 620 |
| HFHT | $4\times10^{16}$ | 5 | 620 |
| LFLT | $4\times10^{15}$ | 0.5 | 253 |
| HFLT | $4\times10^{16}$ | 5 | 253 |

Table II. Saturation magnetization $M_S$ and coercivity $H_C$ determined by SQUID magnetometry. The measurement temperatures are indicated.

| Sample | $M_S$ ($\mu_B$ per implanted Fe) | | $H_C$ (Am$^{-1}$) | |
|---|---|---|---|---|
| | 5 K | 300 K | 5 K | 300 K |
| HFHT | 0.30 | 0.17 | $2.4\times10^4$ | $2.4\times10^3$ |
| LFLT | 1.3 | 1.0 | $4.8\times10^3$ | $4.0\times10^3$ |

**Figure captions:**

Fig 1. Magnetization reversal recorded at 300 K using SQUID magnetometry for the HFHT (a), and the LFLT (b) sample. The inset shows the magnetization prior to background subtraction in $Am^{-1}$ with respect to the substrate volume. The horizontal axes have the same scale. (c) Conventional (Conv.) and SR-XRD pattern (symmetric 2θ/ω scan) for the HFHT sample compared to a virgin sample. Small Fe nanoparticles can be detected only by SR-XRD. (d) SR-XRD pattern for the LFLT sample: no secondary phases are found by either a symmetric 2θ/ω scan or a grazing incidence scan.

Fig. 2. CEMS of the HFHT sample recorded at 300 K. The fit curves represent (from top to bottom) a single emission line corresponding to a $Fe^{3+}$ state (IS=0.53 mm/s with respect to α-Fe), a quadrupole split emission line (QS=0.6 mm/s) corresponding to a $Fe^{2+}$ state (IS=0.69 mm/s), a sextet line resulting from a magnetic hyperfine splitting of a metallic $Fe^0$ state (IS=0.06 mm/s) and a strongly quadrupole split line (QS=1.3 mm/s) of a $Fe^{2+}$ state (IS=0.78 mm/s). The inset shows Moiré contrasts measured using TEM that can be associated with small metallic Fe nanoparticles corresponding to the CEMS results. The arrows indicate the sextet.

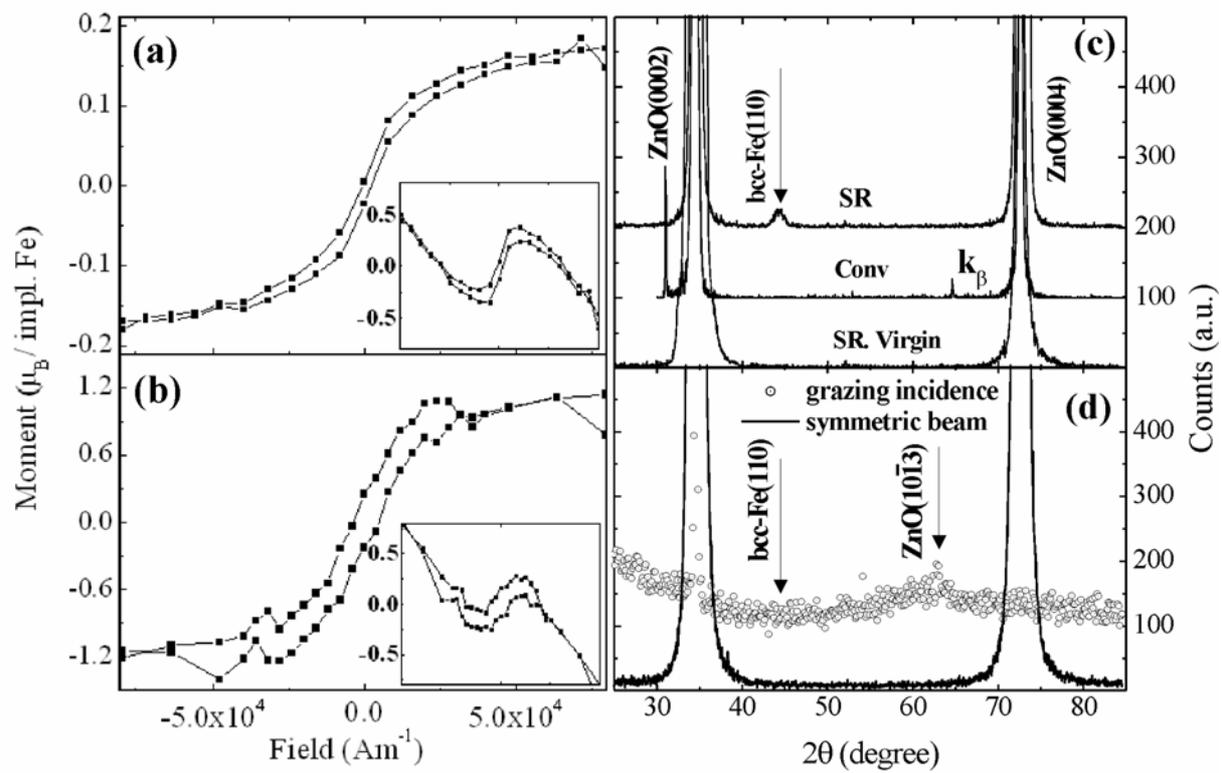

**Fig. 1**

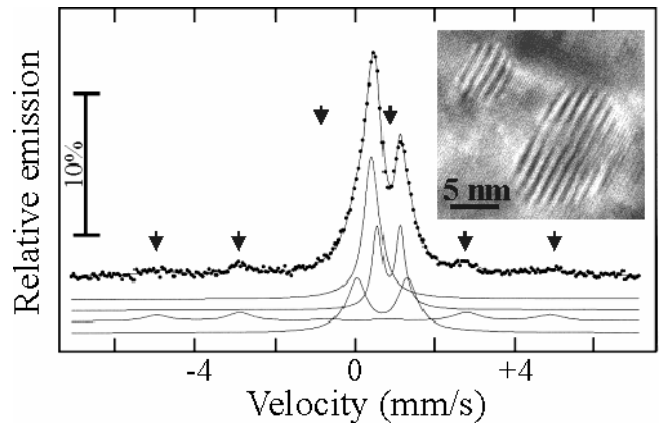

Fig. 2